\documentclass[twocolumn,showpacs,preprintnumbers,amsmath,amssymb]{revtex4}
\usepackage{graphicx}
\newcommand{\dd}{{\rm d}}
\newcommand{\kk}{{\bf k}}
\newcommand{\qq}{{\bf q}}
\newcommand{\rr}{{\bf r}}
\begin{document}
\title{Vibrational excitations in systems with correlated
disorder}
\author{W. Schirmacher}
\affiliation{Phys.-Dept. E13, Technische
Universit\"at M\"unchen, Germany}
\author{B. Schmid}
\affiliation{Fachbereich Physik, Universit\"at Mainz, Germany}
\author{C. Tomaras}
\affiliation{Phys.-Dept. E13, Technische
Universit\"at M\"unchen, Germany}
\author{G. Viliani}
\affiliation{Dipt. di Fisica, Universit\'a di Trento, Italy}
\author{G. Baldi}
\affiliation{INFM-CNR CRS-SOFT OGG, c/o E.S.R.F., Beamline ID16, BP220,
F-38043 Grenoble Cedex, France}
\author{G. Ruocco, T. Scopigno}
\affiliation{ipt. di Fisica, Universit\'a di Roma, Italy;
CRS SOFT-INFM-CNR
c/o Universit\'a di Roma, Italy}

\begin{abstract}
  We investigate a $d$-dimensional model 
($d$ = 2,3) for sound waves in a disordered
  environment, in which the local fluctuations of the elastic modulus
  are spatially correlated with a certain correlation length. The model
  is solved 
analytically by means of a
  field-theoretical effective-medium theory (self-consistent Born approximation) and
numerically on a square lattice. As in the uncorrelated case
the theory predicts an enhancement of the density of states
over Debye's $\omega^{d-1}$ law (``boson peak'')
as a result of disorder. This
anomay becomes reinforced for increasing correlation length $\xi$.
The theory predicts that $\xi$ times the width of the Brillouin line
should be a universal function of $\xi$ times the wavenumber.
Such a scaling is found
in the $2d$ simulation data,
so that they can be represented in a universal plot.
In the low-wavenumber regime, where the
lattice structure is irrelevant there is excellent agreement between
the simulation at small disorder. At larger disorder the continuum
theory deviates from the lattice simulation data. It is argued
that this is due to an instability of the model with stronger
disorder.
\end{abstract}
\maketitle

\section{Introduction~} 
The influence of quenched disorder on the dynamic properties
of solids is enormous and is subject to widespread experimental
and theoretical investigations \cite{binder}.
The disorder leads to strong modifications of physical properties
of the solid. On the other hand,
 the absence of lattice order in the solids and the
corresponding breakdown of the Bloch theorem leads to 
appreciable difficulties
for the theoretical interpretation of the dis\-order-induced phenomena.
Here mean-field theories and particularly effective-medium theories
have been of much help \cite{alexander}, as they very often lead, at least,
to a qualitative understanding of the influence of the disorder.
In particular the coherent-potential approximation \cite{krumhansl}
(CPA)
and its small-disorder version, the self-consistent Born approximation
(SCBA) \cite{ballentine}, have proved to be useful for interpreting the
electronic and other spectral properties of disordered solids.
In many cases physically different situations can be mapped onto
each other. So one can convert an electronical problem to a vibrational
one by replacing the energy $E$ by $-\omega^2$, where $\omega$ is
the frequency parameter. If $E$ is replaced by $i\tilde\omega$, one
studies the mathematical analogous diffusion problem 
\cite{alexander,schirm01,schirm1},
where $\tilde\omega$ denotes the time Fourier parameter
of the diffusion dynamics.

In the case of vibrational properties the disorder-in\-duced excess 
in the density of states (DOS) over 
Debye's $\omega^{d-1}$ law ($d$ is the dimensionality) (``boson peak'')
has been successfully explained for a lattice model by comparing
a simulation of a disordered lattice system
with the predictions of the lattice CPA \cite{schirm0}.
The comparison showed once more
that the CPA is a reliable theory of disorder.
In this study it
was shown that the boson peak anomaly marks the crossover from
plane-wave like vibrational states to disorder-domi\-nated states
with increasing frequency $\omega$. Near the crossover the effective
sound velocity $v$ becomes complex and frequency dependent. This
corresponds to a dc-ac crossover in the analogous diffusion
problem \cite{schirm1}. 

A similar but lattice-inde\-pendent approach
is the generalised elasticity theory in which the disorder
is assumed to lead to spatial fluctuations of the elastic
constants \cite{schirm2}. 
This model was solved by functional-integral techniques
in which the SCBA plays the role of a saddle-point
in a non-linear sigma-model treatment.
This theory allowed for a generalization
to include transverse degrees of freedom and to formulate
a thermal transport theory \cite{schirm3}. Within this theory
it was shown that the so-called plateau in the temperature dependence
of the thermal conductivity \cite{freeman} is caused by the
boson-peak anomaly \cite{schirm3}.
Furthermore, it has turned out \cite{schirm4} that the sound attenuation
parameter (which is proportional to the imaginary part of $v(\omega)$)
is related to the excess in the DOS in the boson peak regime.

It should be noted that the model of spatially fluctuating elastic
constants, treated in CPA and SCBA is by no means the only approach
for explaining the boson-peak anomaly. In fact, an enormous number
of possible explanations have been published in the literature, which
can roughly grouped into three classes: $i$) defect models, $ii$) models
associated with the glass transition and $iii$) models with spatially
fluctuating elastic constants. 

$i$): Defects with a heavy mass can produce
resonant quasi-local resonant
states within the DOS \cite{economou,burin,polishchuk} and be thus
the reason for the boson peak and the reduction of the thermal
conductivity. Similarly defects with very small elastic constants,
near which anharmonic interactions are important (soft potentials),
can produce quasi-local states, which, if hybridized with
acoustic excitations may produce a boson peak \cite{soft,gurevich}
and a plateau in the thermal conductivity
\cite{buchenau,gil}
Inhomongeneities may also be the source of local vibrational
excitations that contribute to the excess DOS \cite{duval}. Specifically
in network glasses bond-angle distortions may also contribute
to the boson-peak anomaly \cite{nucker,nakayama}. In a recent
study \cite{schirm2} the predictions of a defect model
has been compared with those of a fluctuating elastic constant model.
$ii$): In theories of the glass transition
\cite{gotzemayr,chong,xu,lubchenko,angell} the boson peak arises
as a benchmark of the frozen glassy state. $iii$) In models with
quenched disorder of elastic constants 
\cite{schirm01,schirm1,taraskin,%
parisi1,parisi2,bunde,kuhn,schirm2,schirm3,schirm4} the boson peak marks the
lower frequency bound of a band of irregular delocalized states
with random mutual hybridization. These states
are neither propagating nor localized \cite{schirm0}.
The models have been solved 
with the help of
numerical simulations as well as effective-medium theories.

A drawback of these models is that 
they are based on the
model assumption of uncorrelated disorder, i.e. spatial fluctuations
of the physical quantities are assumed to be uncorrelated. This assumption is only justified if the spatial correlation length is smaller
or of the order of the natural length which appears in the system
under consideration. In the present problem, namely
vibrations in disordered solids, there are two important length
scales. One is just the interatomic distance $a$, the other is
the sound velocity, divided by the boson-peak
frequency, which is in experimental data of the order
of several interatomic distances. The wavenumber corresponding to this length scale is the maximum wave\-number
wich can serve as a label for wave-like vibrational states.
In any case
it is a more sound procedure to start with a theory with
{\it correlated disorder} and make the approximation of
short correlations (if appropriate) only in the end. Such
theories are available \cite{john,ignat,bernhard}. The aim
of the present contribution is twofold: First we summarise
the main features of
the long-range-order SCBA. Secondly we
present results of a two-dimensional simulation and compare
them with those of the analytic theory. 
\section{Model and self-consistent Born approximation (SCBA)~}
We start with the equation of motion for 
scalar wave-like excitations $u(\rr,t)$
in a $d$-dimensional disordered medium
\begin{equation}\label{eqmo}
\frac{\partial^2}{\partial t^2}\,u(\rr,t)
=\nabla\cdot\,v^2(\rr)\,\cdot\nabla\,u(\rr,t)
\end{equation}
Here $v(\rr)$ is a sound velocity (and $v^2(\rr)$ an elastic
constant), which is supposed to exhibit random spatial
fluctuations $v^2(\rr)=v_0^2+\Delta(\rr)$ with $v_0^2=
\langle v^2\rangle$ and 
\begin{eqnarray}\label{corr}
&&\langle\Delta(\rr+\rr_0)\Delta(\rr_0)\rangle
=C(\rr)=\langle\Delta^2\rangle e^{-r/\xi}\nonumber\\
\Leftrightarrow\quad&&\qquad C(\kk)=\langle\Delta^2\rangle C_0\,
[k^2+\xi^{-2}]^{-\frac{d+1}{2}}
\end{eqnarray}
with $C_0=2\pi(3d\!-\!5)/\xi$.
In an effective-medium approximation the disordered system
is mapped onto a homogeneous system, in which the influence
of the disorder enters via a self-energy function
$\Sigma(\kk,z)$ with $z=\omega+i\epsilon$. The dynamic
susceptibility is given by
\begin{equation}
\chi(\kk,z)=\frac{k^2}{-z^2+k^2(v_0^2-\Sigma(\kk,z))}
\end{equation}
In SCBA the self-energy function obeys the self-consistent
equation
\cite{ballentine,john,ignat,bernhard}
\begin{equation}\label{scba}
\Sigma(\qq,z)=\frac{1}{2}\int\limits_{|\kk|<k_D}
\left(\frac{\rm d\kk}{2\pi}\right)^d
C(\qq-\kk)\chi(\kk,z)
\end{equation}
with the
Debye wavenumber
\mbox{$k_D=[2d\pi^{d-2}]^{1/2}/a$}.

In the present study we are mainly interested in the low-frequency
and -wavenumber properties, so we replace the self-energy
by its $q\rightarrow 0$ limit $\Sigma(z)$ and obtain the SCBA
equation
\begin{equation}\label{scba3}
\!\Sigma(z)=\frac{\gamma}{2}
\,\,{\textstyle 
\frac{v_0^4}{\varphi_d}
}
\!\!\!\int\limits_{|\kk|<k_D}\!\!\!\!\left(\frac{\dd \kk}{2\pi}\right)^d
\!\frac{k^2C(k)/\langle\Delta^2\rangle}{
-z^2+k^2\left(v_0^2-\Sigma(z)\right)}
\end{equation}
with the normalization constant
$\varphi_d={\scriptstyle \!\!\!\!\int\limits_{|\kk|<k_D}\!\!\!\!(\dd \kk/2\pi)^d
C(\kk)/\langle\Delta^2\rangle}$ and the ``disorder parameter''
$\,\,\gamma=\langle \Delta^2\rangle\varphi_d/v_0^4$.\\

\begin{figure}[t]
\centerline{\includegraphics*[width=\linewidth,height=6cm]{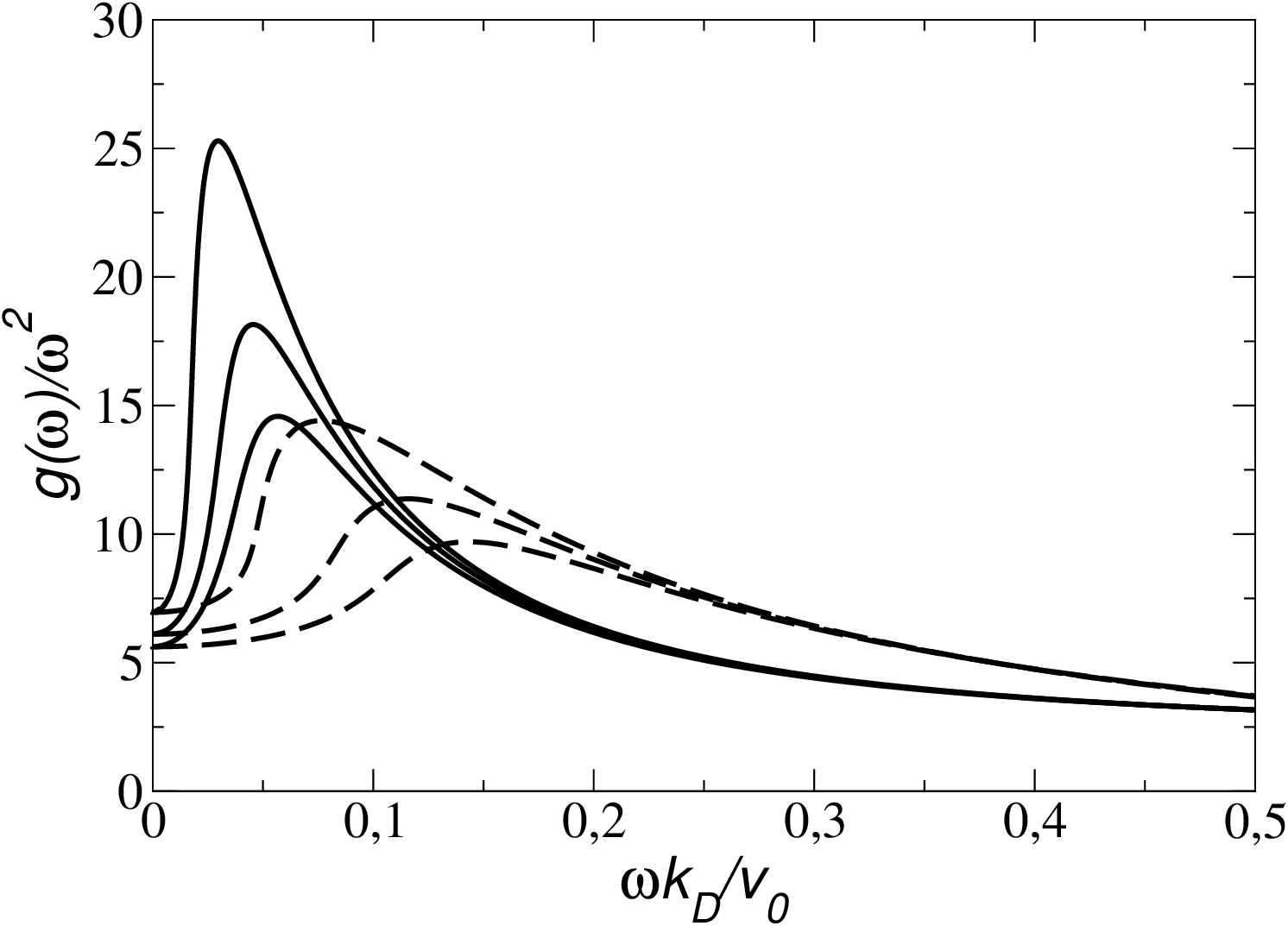}}
\caption{Reduced DOS $g(\omega)/\omega^2$ for
$\xi\!=\!1/k_D$ (dashed lines) and $\xi\!=\!5/k_D$ (full lines) and
for three disorder parameters (from left to right)
$\gamma$ = 0.49, 0.47 and 0.45.}
\end{figure}
The DOS is given by
\begin{equation}\label{dos}
g(\omega)=\frac{2\omega}{\pi}{\rm Im}\left\{\int\left(\!\frac{\rm d\kk}{2\pi}\right)^d
\!\!\frac{1}{-z^2+k^2(v_0^2-\Sigma(z))}\right\}
\end{equation}
In Fig. 1 we have plotted the ``reduced DOS''
$g(\omega)/\omega^2$ for $d=3$ and
for three values of $\gamma$ and
two values of $\xi$. First we notice that, as in the
uncorrelated case \cite{schirm0,schirm2,schirm3} there
exists a critical amount of disorder $\gamma_c=0.5$, beyond
which the system becomes unstable. The ``boson peak'' becomes
more pronounced and situated at lower frequencies as this
value is approached. The interesting new feature in the correlated
case is that the boson peak is re-inforced by the correlation
and again shifted towards lower frequencies.
\section{Dynamic structure factor and density of states}
We now divide the self-energy function into a real
and imaginary part 
$\Sigma(z)\!=\!\Sigma'(\omega\!\approx \! 0)\!+\!i\Sigma''(\omega)$ and
define the {\it renormalised} sound velocity as
\mbox{$v^2=v_0^2\!-\!\Sigma'$.}
The dynamical structure factor $S(\kk,\omega)$, which can be
measured by inelastic neutron or X-ray scattering,
and which is the Fourier transform of the
dynamic density-density correlation function,
is then given by the fluctuation-dissipation theorem
\cite{hansen} as
\begin{eqnarray}
\qquad&&\qquad S(k,\omega)=\frac{k_BT}{\pi\omega}
\mbox{Im}\left\{
\chi(k,z)
\right\}\nonumber\\
\qquad&&\approx
\frac{k_BT}{2\pi v^2}\frac{k^2\Sigma''(\omega)/2\omega}
{\left[(kv-\omega)^2+(k^2\Sigma''(\omega)/2\omega)^2\right]}
\end{eqnarray}
The Brillouin resonance is given by $\omega=kv$ and the
line width (FWHM, sound attenuation parameter)
is given by
\begin{figure}
\centerline{\includegraphics*[width=\linewidth,height=6cm]{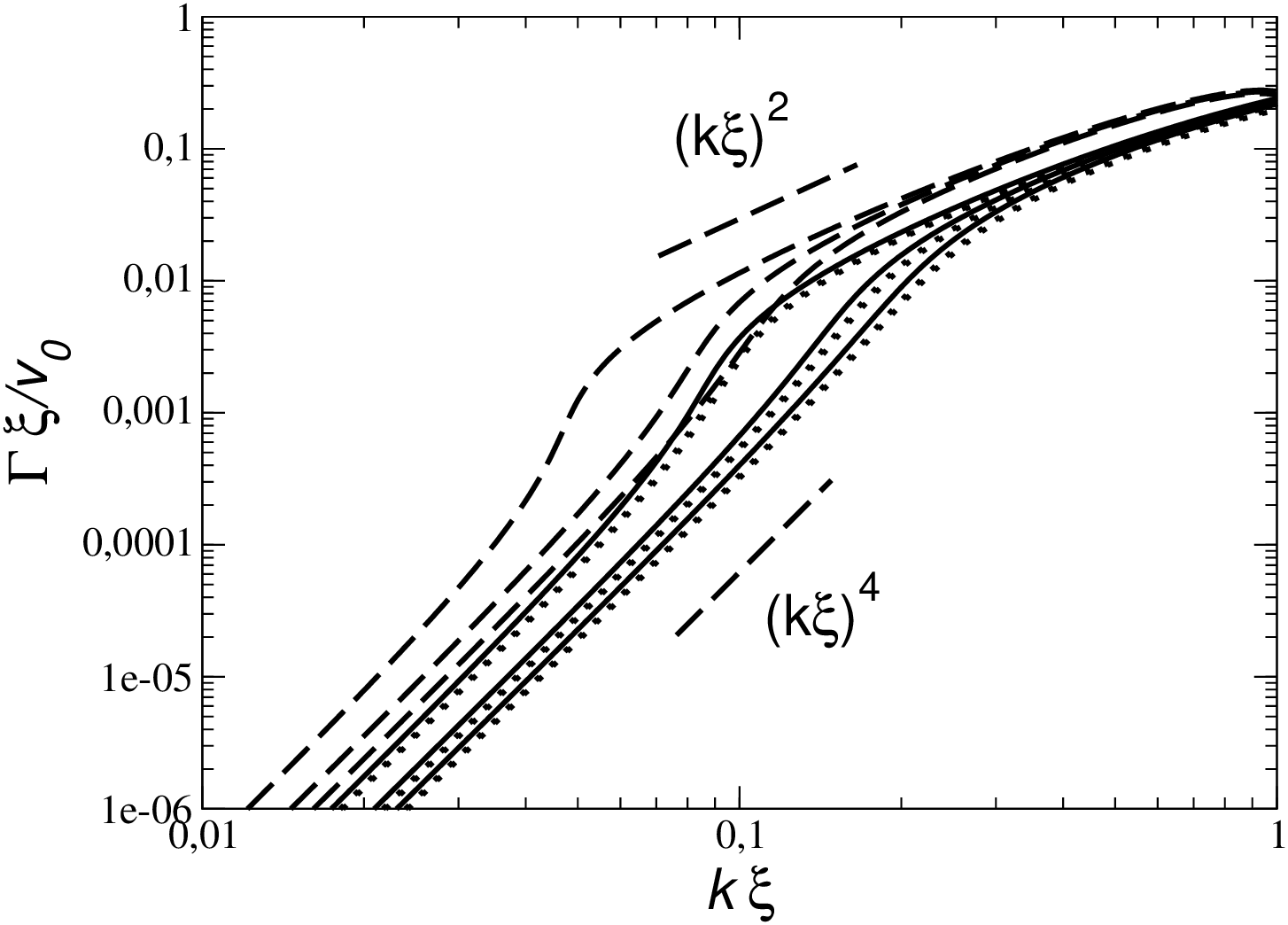}}
\caption{%
Scaled Brillouin linewidth $\Gamma\xi/v_0$ against scaled
wavenumber $q\xi$ for 
$\xi\!=\!1/k_D$ (dashed lines) \cite{comment},
$\xi\!=\!5/k_D$ (full lines), 
$\xi\!=\!10/k_D$ and
$15/k_D$ (dotted lines). The disorder parameters
$\gamma$ are the same as in Fig. 1.
}
\end{figure}
\begin{equation}\label{width}
\Gamma(k)=\frac{k}{v}\Sigma''(\omega=v\,k)
\end{equation}
It can easily be shown that $\Sigma''\propto \omega^d$ for
$\omega\rightarrow 0$ so that $\Gamma\propto q^{d+1}$ for
$q\rightarrow 0$ (Rayleigh law).

We~~~now~~~introduce~~~the~~~dimensionless~~~variables
\mbox{$\tilde \Sigma\!=\!\Sigma/v_0^2\,$},
\mbox{$\,\,\tilde C_0=C_0\xi=2\pi(3d-5)$},
\mbox{$\,\,\tilde z\!=\!z\xi/v_0
\!=\!\tilde\omega\!+\!i\tilde\epsilon$}\newline and
\mbox{$\tilde k\!=\!k\xi$.}
We then obtain
\begin{equation}\label{scba1}
\tilde\Sigma(\tilde z)=\frac{\gamma}{2}
\,\,{\textstyle\,\frac{\tilde C_0}{\varphi_d}}
\!\!\!\!\int\limits_{|\tilde \kk|<\xi k_D}
\!\!\!\left(\frac{\rm d\tilde\kk}{2\pi}\right)^d
\!\frac{\!\tilde k^2\,[1+\tilde k^2]^{-(d+1)/2}}
{-\tilde z^2+\tilde k^2(1-\tilde \Sigma(\tilde z))}
\end{equation}
In this expression the correlation length $\xi$
enters only
via the upper $\tilde k$ cutoff.
For the limit $\xi\gg a$ we therefore
obtain the scaling
relation
\begin{equation}
\tilde \Gamma=\Gamma\xi/v_0=f(\gamma,\tilde k)
\end{equation}
In Fig. 2 we have plotted $\tilde \Gamma$ against
$\tilde k$ for $d=3$, for the three $\gamma$ values
of Fig. 1 and for $\xi$ = 
$1/k_D$,
$5/k_D$,
$10/k_D$, and
$15/k_D$.
We see that the scaling is obeyed except for $\xi=1/k_D$ 
\cite{comment} as
expected. As in the uncorrelated case \cite{schirm4},
the boson peak (see Fig. 1) marks the crossover from
the Rayleigh regime $\tilde \Gamma\propto\tilde k^4$
to a behaviour $\tilde\Gamma\propto\tilde k^s$
with $s\approx 2$. 
\section{Simulation}
We now discretize (\ref{eqmo}) in $d=2$ on a square lattice,
which then takes the form of an equation of motion
for unit masses connected by springs with
spring constants $K_{ij}=\frac{1}{2a^2}(v^2(r_i)+v^2(r_j))$:
\begin{equation}\label{eqmo1}
\frac{\dd^2}{\dd t^2}\,u(\rr_i,t)
=\sum\limits_{\ell \,\,n. N.}K_{i\ell}[u(\rr_{\ell},t)-u(\rr_i,t)]
\end{equation}
\begin{figure*}[htb]%
\parbox{18cm}{\includegraphics*[width=8cm]{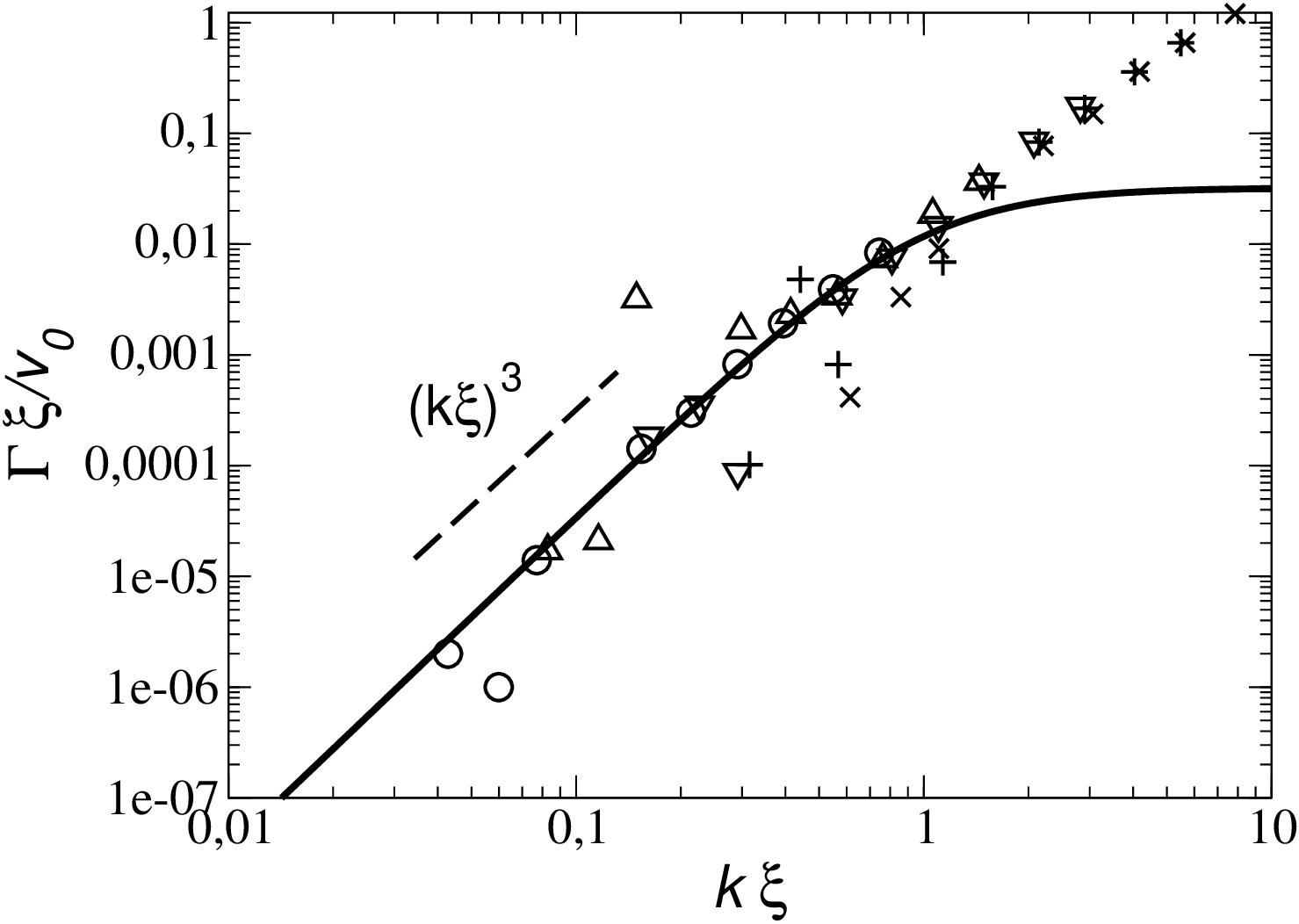}
\includegraphics*[width=8cm]{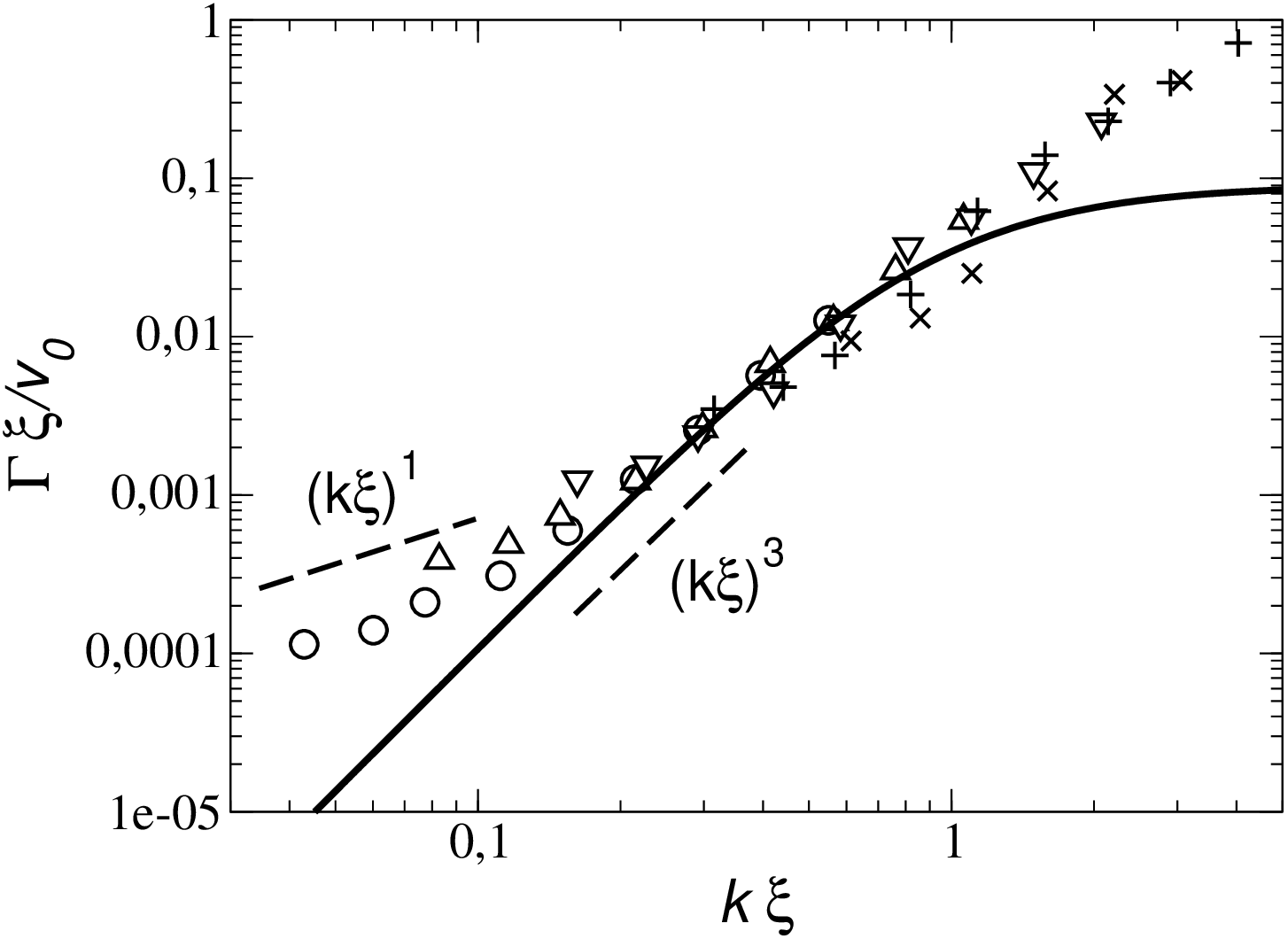}}
\caption{Symbols: Linewidths $\Gamma$ of the simulated Brillouin
spectra, multiplied with the correlation length $\xi$ and
divided by the sound velocity $v$. 
$\,\,\,\circ : \xi=2.79a$,
$\Delta : \xi=5.42a$,
$\nabla : \xi=10.56a$,
{\sf +} : $\xi=20.55a$,
{\sf x} : $\xi=40.0a$,
Straight line: theory.
The results of the left picture correspond to a disorder
parameter $\gamma=0.04$ that of the right picture
to $\gamma=0.111$.}
\end{figure*}
In the simulation \cite{baldi} the force constants
$K_{ij}$ are extracted from a random distribution with mean
$K_0$ and variance $\sigma^2$. The correlation is established
following the Fourier filtering method (FFM) 
\cite{makse}. The network of random springs is created
starting from a random set of $(2N)^2$ numbers uniformly
distributed around zero obtained from a pseudo random
number generator. The FFM method is then used to generate
a $2N\times 2N$ two-dimensional lattice of ``pair'' random
numbers $\{\eta_{ij}\}$ which obey a spatial
correlation
\begin{equation}
\left\langle \eta_{00}\eta_{ij}\right\rangle
\propto e^{-\frac{a}{2}(i+j)/\xi}
\end{equation}
The random spring constants with mean $K_0$ and variance
$\sigma^2$ are extracted from the pair numbers $\eta_{ij}$.
The lattice spacing of masses-and-springs system
is twice of that of the random-number lattice. 
Using this statistics the dynamic structure factor 
$S(k,\omega)$
of the
model has been determined by the method of moments
\cite{benoit1,benoit2,viliani}.
In Figs. 3a and 3b the scaled widths of the Brillouin peak
of samples with different correlation lengths have been plotted
against the scaled wavenumber $\tilde k$. It is clearly seen
that the simulated data follow the predicted scaling law.
For the case with the lesser disorder ($\gamma
=(\sigma/K_0)^2=0.04$) there
is very good agreement with the theory (~(\ref{scba1}) with
no cutoff in the integral) in the small wavenumber limit,
where the lattice and continuum models should agree.
In the high-wavenumber regime, of course, the lattice
character of the simulated system becomes distinct.

Let us turn to the discussion of the data of Fig. 3b
with the increased disorder $\gamma=0.111$. The continuum
theory in this case predicts the 
Rayleigh law $\tilde \Gamma\propto \tilde k^3$
(continuous line). 
We checked the stability of the system by investigating
the 
simulated
density of levels $g(\omega^2)=g(\omega)/2\omega$ and
found that this quantity exhibits nonzero values for
$\omega^2<0$, which means that the system is unstable.
For this case it is known that the imaginary part
of the self energy $\Sigma''$ is constant and passes continuously
from positive to negative values of $\omega^2$ in this case.
Consequently the line width $\Gamma(\omega)\propto k\Sigma''$ 
shows a linear increase for small omega.
Such a behavior is obviously
an artefact of constructing a harmonic model with too much
disorder, which leads to a small fraction
of negative elastic constants. In a realistic physical system
such ``would-be'' negative elastic constants are removed by
the anharmonic interaction which causes relaxation of the
system towards a stable situation. This has been nicely demonstratedby a model calculation of Gurevich et al \cite{gurevich}.
\section{Conclusion}
We~ have~ investigated~ the vibrational properties of disordered
systems with correlated disorder both analytically by the
self-consistent Born approximation as well as by a simulation
applying the method of moments
and the Fourier-filtering method. The sound attenuation
constant $\Gamma(k)$
(width of the Brillouin line) is found to scale
with the correlation length $\xi$ in such a way that
$\xi\Gamma$ is a universal function of $\xi k$. The
enhancement of the density of states (boson peak) is found
to be re-inforced by the correlations. 

\section*{Ackowledgement:}~~~W. S. is grateful for hospitality
at the Universit'a di Roma, ``la Sapienza'' and at
the University of Mainz.

\end{document}